\documentclass[conference]{IEEEtran}
\usepackage{cite}
\usepackage{amsmath,amssymb,amsfonts}
\usepackage{algorithmic}
\usepackage{graphicx}
\usepackage{textcomp}
\usepackage{booktabs} 
\usepackage{multirow}
\usepackage{bm}
\usepackage{threeparttable}

\def\BibTeX{{\rm B\kern-.05em{\sc i\kern-.025em b}\kern-.08em
		T\kern-.1667em\lower.7ex\hbox{E}\kern-.125emX}}

\usepackage{hyperref}

\begin{document}
	
	\title{Enhancing Community Detection in Networks: A Comparative Analysis of Local Metrics and Hierarchical Algorithms}
	
	\author{
		\IEEEauthorblockN{Julio-Omar Palacio-Niño}
		\IEEEauthorblockA{Dept. Computer Science and\\
			Artificial Intelligence\\
			University of Granada\\
			Granada, Spain\\
			Email: jopalacion@correo.ugr.es}
		\and
		\IEEEauthorblockN{Fernando Berzal}
		\IEEEauthorblockA{Dept. Computer Science and\\
			Artificial Intelligence\\
			University of Granada\\
			Granada, Spain\\
			Email: berzal@acm.org}
	}
	
	\maketitle

\begin{abstract}
	The analysis and detection of communities in network structures are becoming increasingly relevant for understanding social behavior. One of the principal challenges in this field is the complexity of existing algorithms. The Girvan-Newman algorithm, which uses the betweenness metric as a measure of node similarity, is one of the most representative algorithms in this area. This study employs the same method to evaluate the relevance of using local similarity metrics for community detection. A series of local metrics were tested on a set of networks constructed using the Girvan-Newman basic algorithm. The efficacy of these metrics was evaluated by applying the base algorithm to several real networks with varying community sizes, using modularity and NMI. The results indicate that approaches based on local similarity metrics have significant potential for community detection.
\end{abstract}

\section{Introduction}
\label{sec:introduction}
Networks are ubiquitous, making their analysis essential across various scientific fields such as molecular biology, sociology, epidemiology, and computer science. Community detection, defined an unsupervised machine learning task, is pivotal within the realm of network data mining. Although numerous techniques have been proposed to detect communities within complex networks\cite{1},achieving efficiency remains a significant challenge for these algorithms\cite{2}.
\\
Among the most successful approaches is hierarchical clustering for community detection. The Newman and Girvan algorithm \cite{3} is a notable hierarchical divisive method that utilizes the betweenness of edges to discern links acting as bridges between distinct communities. Unfortunately, this approach requires the recalculated of edge betweenness each time a link is removed, entailing a computationally demanding process. Radicchi \cite{4} proposed an alternative algorithm that leverages the clustering coefficient for local calculations, thereby enhancing efficiency.
\\

Alongside the evolution of unsupervised community detection algorithms, research has also advanced in supervised domains such as link prediction \cite{5}. Link prediction algorithms aim to forecast the existence or future emergence of links between node pairs, relying on structural network properties, both local and global. Given the limitations inherent to local properties, their application has extended to developing more efficient algorithms for both link prediction and community detection.
\\
This paper examines the potential for adapting local structural properties, originally designed for supervised learning tasks, to unsupervised learning in order to enhance the efficacy of hierarchical community detection algorithms. The article is structured as follows: Section 2 discusses hierarchical methods for community detection. Section 3 focuses on the problem of link prediction, emphasizing local techniques and relevant structural properties. Section 4
presents experimental results using reference network datasets. Section 5 concludes with insights and recommendations on the utilization of local structural properties for community detection.
\\

\section{Related work}

One of the most effective methods for representing related information is through the use of networks or graphs. In this context, objects are referred to as nodes, which are connected by edges \cite{6} . One of the earliest modeling approaches involves unweighted and directed graphs. An undirected graph is defined as a pair $G=(V,E)$ ,where $V$ is a finite set of nodes and $E$ is a set of unordered pairs of elements. This set of unordered pairs is known as edges \cite{7}. The problem of community discovery is not clearly defined; one approach to addressing it relies on unsupervised learning for cluster discovery \cite{1}.

\subsection{Detection of Hierarchical Communities}

Hierarchical clustering is an unsupervised technique of machine learning that iteratively runs until the formation of a dendrogram, where each leaf represents individual instances and the root encompasses the entire dataset \cite{1}.
\\
Hierarchical algorithms operate from two perspectives: agglomerative and divisive. Agglomerative algorithms traverse the information from the leaves to the root, generating clusters. Conversely, divisive algorithms start from the root and progress through the tree to its leaves, forming groups by dividing them based on their differences. Due to the characteristics and size of the data, it is more common to begin with a large dataset and progressively divide it until a small number of clusters are formed; therefore, divisive methods are usually preferred.
\\
\subsection{Girvan-Newman Algorithm}

Girvan and Newman proposed a hierarchical divisive algorithm for the detection of communities in networks, using "edge betweenness" as the key metric to identify links that separate communities. Edge betweenness is a fundamental concept in network analysis, defined as the count of the shortest paths that pass through each edge.
\\
\\
The greater the number of these paths, the higher the likelihood that an edge connects different communities. This metric is instrumental in determining the sequence of divisions in a traditional hierarchical divisive algorithm, guiding the process of delineating community boundaries effectively.
\\
\\
Formally, the "Edge betweenness" for each edge is denoted as \( e \in E \).
\\
\\
\[
C_B(e) = \sum_{u, v \in V} \frac{g_e(u, v)}{g(u, v)}
\]
\\
\\
where $g(u,v)$ is the minimum number of edges between the nodes $u$ and $v$, is the minimum number of edges between and but passing through the node $e$ . Wherefore the betweenness coefficient for the node $e$ is denoted as $C_B(e)$. Small values of $C_B(e)$ indicate that the edges belong to the same community, while those edges that connect different communities as "bridges" will have higher values. \cite{8}.
\\
To execute a complete division process, Girvan and Newman's algorithm computes the betweenness coefficient for each node in the network. Then, iteratively, as long as an edge's betweenness remains above a predefined threshold, the edge with the highest value is removed, and the coefficient is recalculated for the remaining network.
\\
One of the principal drawbacks of Girvan and Newman’s algorithm is its algorithmic complexity. In each iteration, the betweenness coefficient of every edge must be recalculated, resulting in a worst-case complexity of $O(n3)$ in highly dense networks, where $m \propto n$ indicates the number of edges m is proportional to the number of nodes n. \cite{8} Despite yielding favorable results, this complexity constrains its practical application in large networks.
\\

\subsection{Radicchi algorithm}

Radicchi's algorithm \cite{4}, similar to that of Girvan and Newman, aims to identify communities within a network. However, unlike the latter which relies on calculating an edge betweenness coefficient, Radicchi's method is based on the concepts of connectivity and clustering.
\\
\\
A community is characterized by highly connected nodes, leading to frequent cycles among nodes within the same community. Conversely, links that connect different communities tend to participate in fewer cycles, thus having a lower probability of being part of a cycle \cite{9}.
\\
\\
Radicchi's algorithm employs an edge clustering coefficient to determine an edge's likelihood of belonging to a community, which is formally defined as:
\\
\\
\[
c_{ij} = \frac{z_{ij} + 1}{\min(k_i - 1, k_j - 1)}
\]
\\
\\
Where $z_{ij}$ is the number of triangles formed with that edge, involving nodes $i$ and $j$, $k_i$ is the degree of node $i$, and $k_j$ the degree of node $j$. The numerator +1 added to $z_{ij}$ prevents excessively penalizing edges that do not belong to any triangles. In instances where $k_i$ or $k_j$ is 1, the equation would diverge; hence, such edges are excluded from consideration. In the Radicchi algorithm, smaller coefficients $c_{ij}$ indicate membership within a community. The algorithm involves removing edges with low coefficients and then recalculating the coefficient for the remaining edges \cite{9}.
\\
\\
Radicchi's algorithm specifically targets small cycles of length 3, recognizing that edges facilitating inter-community connections typically exhibit a low clustering coefficient. This highlights the clustering-coefficient as a reliable indicator for identifying bridges between communities \cite{4}.

\section{Local clustering coefficient}

One of the most commonly used metrics for evaluating the quality of communities is the clustering coefficient. This measure assesses the degree to which nodes in a network belong to a cluster. In Radicchi's algorithm, the clustering coefficient helps identify triangles, indicating nodes that are highly connected, similar to the closeness observed among friends in a social network \cite{10}.
\\
\\
\[
C_i = \frac{2 m_i }{k_i(k_i - 1)}
\]
\\
\\
Where $m_i$ is the total number of connections between $k_i$ neighbors of node $i$. Its interval is contained between [0,1]. $C_i=0$ will indicate that the neighbors of node $i$ are not connected to each other at all, while a $C_i=1$ will indicate that all the neighbors of node i are connected to each other. It is also possible to represent the global clustering coefficient from the local measures\cite{11}.
\\
\\
\[
C_g = \frac{1}{n} \sum_{i \in V} C_i
\]
\\
\\
\subsection{Local link prediction}
Link prediction in network structures is a supervised learning approach that can improve the efficiency of the Girvan and Newman algorithm and address the limitations of the Radicchi algorithm. This technique aims to leverage the intrinsic properties of nodes and edges, as well as their behaviors within the network, to enhance community detection \cite{8}.
\\
\\
Employing a hierarchical approach, the process iteratively decides which edges should be eliminated or retained to construct subgraphs \cite{12}. Girvan and Newman's algorithm uses the modularity criterion to evaluate the strength of each edge in relation to the overall graph. Similarly, Radicchi's algorithm, based on the same principle, aims to establish connectivity between nodes through the connectivity principle.
\\
\\
Over the years, a few techniques have been proposed for the prediction of links in complex networks. These include local medical, quasi-local, classification-based, metaheuristic, and factor-based techniques \cite{5}.
\\
\\
Research conducted by Liben-Nowell and Kleinberg \cite{13}  , Zhou et al. \cite{14}, and others have demonstrated that local link prediction methods produce favorable outcomes with significantly less computational complexity compared to global clustering-based techniques \cite{15}. Similarly, studies by Wu \cite{16} and Kumar \cite{17} have explored link prediction strategies that rely on the clustering coefficient.
\\
\\
It is important to note that local link prediction methods inherently cannot predict the existence of links between remotely located nodes. Nonetheless, because most links tend to form within a specific neighborhood, and considering the computational advantages they provide, local link prediction techniques are ideally suited to enhance the performance of existing hierarchical community detection methods.
\\
\\
Among the most common measures for link prediction are:
\\
\\
\textbf{Common Neighbors (CN):} For a node $x$, $\Gamma_x$ denotes the set of communities to which $x$ belongs. This measure is grounded in identifying the number of neighbors common between a pair of nodes $(x,y)$. The greater the number of common neighbors, the higher the probability of an edge existing between them. The similarity function is defined as \cite{18}:
\\
\\
\[
s(x, y) = |\Gamma_x \cap \Gamma_y|
\]
\\
\\
Values close to zero suggest that the two nodes have no neighbors in common and are therefore "farther away" from each other. Conversely, a value close to one indicates that the nodes are "close" to each other. Due to its simplicity, this measure forms the foundation for other link prediction and evaluation metrics. Its computational complexity is estimated to be $O(vk^3)$ where v represents the number of nodes and k the maximum degree of nodes \cite{5}.
\\
\\
\textbf{The Adamic-Adar Index (AA):} This metric evaluates the connection strength between a pair of nodes based on the number of common edges they share, similar to the Common Neighbors approach. However, the Adamic-Adar Index enhances this basic measure by weighting the shared edges according to the inverse of their frequency of occurrence. This approach recognizes that not all shared connections contribute equally to the likelihood of a link, with less frequent connections often indicating stronger mutual relevance\cite{19}:
\\
\\
\[
s(x, y) = \sum_{z \in \Gamma_x \cap \Gamma_y} \frac{1}{\log |\Gamma_z|}
\]
\\
\\
\textbf{The Resource Allocation Index (RA):}This metric is based on the process of allocating resources from one community to another through a common path that includes two shared nodes $(x, y)$ \cite{14}:
\\
\\
\[
s(x, y) = \sum_{z \in \Gamma_x \cap \Gamma_y} \frac{1}{|\Gamma_z|}
\]
\\
\\
\textbf{The Preferential Attachment Index (PA):}This index considers that the existence of a link between two nodes is determined simply by the number of links each node has. In this way, the presence of an edge between a pair of nodes is determined by the number of connected nodes, without considering any additional information \cite{20}:
\\
\\
\[
s(x, y) = |\Gamma_x| \cdot |\Gamma_y|
\]
\\
\\
\textbf{The Jaccard Index (JA):}Proposed by Paul Jaccard, this index is primarily used to compare the similarity and difference between two sets by analyzing the relationship between a pair of nodes $(x, y)$\cite{21}:
\\
\\
\[
s(x, y) = \frac{|\Gamma_x \cap \Gamma_y|}{|\Gamma_x \cup \Gamma_y|}
\]
\\
\\
\textbf{Sørensen Index (JA):}is a proposal by Sørensen to assess the similarity between ecological data samples \cite{22}. It's also very similar to Jaccard's measurement.
\\
\\
\[+
s(x, y) = \frac{2 |\Gamma_x \cap \Gamma_y|}{|\Gamma_x| + |\Gamma_y|}
\]
\\
\\
\textbf{Salton Index:}Also known as the Salton Cosine Index \cite{23}, it is used to determine the similarity between two datasets based on the cosine similarity between the rows of the adjacency matrix corresponding to nodes x and y \cite{24}.
\\
\\
\[
s(x, y) = \frac{|\Gamma_x \cap \Gamma_y|}{\sqrt{|\Gamma_x| \cdot |\Gamma_y|}}
\]
\\
\\
\textbf{Hub Depressed index:}the depressed hub is the result of the modularity study in metabolic networks conducted by Ravasz et al. \cite{25} it is defined as the proportion of common neighbors between x and y with the highest degree between them.
\\
\\
\[
s(x, y) = \frac{|\Gamma_x \cap \Gamma_y|}{\max(|\Gamma_x|, |\Gamma_y|)}
\]
\\
\\
\textbf{Hub promoted index:}Reciprocally, the Hub Promoted Index is defined as the proportion of common neighbors between X and Y with the minimum degree between them.  \cite{25} 
\\
\\
\[
s(x, y) = \frac{|\Gamma_x \cap \Gamma_y|}{\min(|\Gamma_x|, |\Gamma_y|)}
\\
\\
\]
\textbf{Local Leicht-Holme-Newman Index:}Similar to the Jaccard and Salton indices but more sensitive to differences \cite{26}, it is defined as the ratio of the common neighbors of nodes x and y to the product of the degrees of nodes x and y.
\\
\\
\[
s(x, y) = \frac{|\Gamma_x \cap \Gamma_y|}{|\Gamma_x| \cdot |\Gamma_y|}
\]
\\
\\
It is evident that most local prediction measures are based on the measure of common neighbors, except for the measure of preferential attachment \cite{15}. For more information on these measures, see the works by Lü and Zhou \cite{18} or Martinez et al. \cite{5}, which provide an extensive overview of different link prediction measures.
\\
\\
\subsection{Evaluation metrics}
Determining the exact division of a graph to separate it into two or more communities is not trivial. Various methods have been developed to quantify the quality of the partitioning depending on the characteristics of the graph. Among the most relevant are the Modularity Index and the Normalized Mutual Information Index.
\\
\\
The Modularity Index \cite{8} is one of the most commonly used metrics for community detection, and it is the basis for Girvan-Newman's algorithm. This index compares the set of links within a cluster to what would be expected if the links were distributed randomly. Densities inside and outside each cluster are calculated and compared to the overall density that the network would have if it were random \cite{27}. It is defined as:
\\
\\
\[
Q = \sum_r (e_{rr} - a_r^2)
\]
\\
\\
Where are set of edges within a cluster represents the set of nodes that are part of the cluster \cite{27}. The value of this metric tends to 1; values close to 1 indicate that the cluster is highly
connected. In practice, it has been observed that communities typically present a modularity value between 0.3 and 0 \cite{27}.
\\
\\
Another metric is based on information theory, where two partitions are similar if little information is needed to infer that one partition is equal to another. This amount of information can be measured as entropy \cite{19}.
\\
\\
\[
H(C) = -\sum_{i=1}^k \left[ P(i) \log_2 P(i) \right], \quad P(i) = \frac{|C_i|}{n}, \quad C_i \in C
\]
\\
\\
where for each cluster the probability is calculated. The lower the entropy, the more stable the clustering \cite{1}.
\\
\\
The mutual information index assesses how much the uncertainty of a random node about a cluster can be reduced if the state of the original node and its true cluster membership are known in advance \cite{28}. It is defined as follows:
\\
\\
\[
J(X, Y) = \sum_{i=1}^{n_x} \sum_{j=1}^{n_y} \left[ P(i, j) \log_2 \left( \frac{P(i, j)}{P(i) P(j)} \right) \right] 
\]
\\
\\
\[
\quad P(i, j) = \frac{|X_i \cap Y_j|}{n}
\]
\\
\\
The probability that a node belongs to a cluster and also belongs to another cluster is considered. To limit the value that this measurement can take, normalization is implemented \cite{28}.
\\
\\
\[
\text{"NMI"}_2(X, Y) = \frac{2J(X, Y)}{H(X) + H(Y)}
\]
\\
\\
The important aspect of this measure is its ability to analyze clusters without needing to know the clustering method or the original characteristics of the network directly. The measurement ranges between 0 and 1, where 1 indicates that the partitions are identical, and 0 indicates that the partitions are completely independent \cite{29}.
\section{Framework }
To evaluate how well community detection works using a local metric, we developed a workflow based on Radicchi's algorithm. This workflow includes the following steps:
\begin{itemize}
    \item Calculate the evaluation index for each edge in the graph.
    \item Create a ranked list of these indices for all the edges in the network.
    \item Remove the edge with the lowest value, indicating the weakest link between a pair of nodes. If this splits the graph into two parts, the process stops. Otherwise, repeat the process iteratively until the desired number of communities is reached.
 \end{itemize}
We use two indices for evaluation: modularity and normalized mutual information. Modularity helps determine the optimal number of communities, while normalized mutual information assesses how similar the algorithm's result is to the actual community structure.
\\
\\
Additionally, we use the Girvan-Newman algorithm as a reference point to compare and evaluate the results of our local metric-based algorithm.
\\

\section{experimentation}
The results obtained by the hierarchical community detection algorithm will be analyzed using a dozen different criteria across half a dozen reference datasets. Initially, the Girvan-Newman algorithm based on the betweenness index and the Radicchi algorithm based on the clustering coefficient will be used as benchmarks. Subsequently, the performance of 10 link prediction metrics described in the previous section will be evaluated. The following metrics will be assessed: Common Neighbors (CN), Adamic-Adar Index (AA), Resource Allocation Index (RA), Preferential Attachment Index (PA), Hub Depressed Index (HD), Hub Promoted Index (HP), Jaccard Index (JA), Local Leicht-Holme-Newman (LLHN) Index, Salton Index (SA), and Sørensen Index (SO).
\\
\\
\subsection{Benchmark Dataset}
The following network data sets were obtained from networkrepository.com \cite{30}, SNAP \cite{31}, and Pajek datasets \cite{32}:

\begin{itemize}
    \item Zachary's karate club \cite{33}
    \item Word adjacencies \cite{8}
    \item Dolphin \cite{34}
    \item Les Misérables \cite{35}
    \item U.S. politics \cite{36}
    \item American football \cite{4}
\end{itemize}

\subsection{Modularity Based}
We used a total of 12 different criteria to build the hierarchical community detection algorithm, evaluating its performance in terms of modularity. In optimization-based community detection algorithms, modularity is any numerical measure that helps identify communities in networks. However, in this case, we are specifically referring to the original Q modularity defined by Newman \cite{37}.
\\
\\
Calculating modularity doesn't depend on knowing the number of communities beforehand. We conducted an exploratory analysis ranging from 2 to 10 communities to see how the metric behaves in comparison to the Girvan-Newman reference algorithm.
\\
\\
\begin{table*}[ht]
\centering
\begin{threeparttable}
\caption{Features table}
\label{tab:my-table}
\begin{tabular}{@{}|l|r|r|r|r|r|r|r|r|@{}}
\toprule
\multirow{2}{*}{Networks} & \multicolumn{1}{l|}{\multirow{2}{*}{Nodes}} & \multicolumn{1}{l|}{\multirow{2}{*}{Edges}} & \multicolumn{1}{l|}{\multirow{2}{*}{Deg. Avg.}} & \multicolumn{1}{l|}{\multirow{2}{*}{Eigenvector   CA}} & \multicolumn{1}{l|}{\multirow{2}{*}{Closeness CA}} & \multicolumn{1}{l|}{\multirow{2}{*}{Clustering Coef. Avg}} & \multicolumn{1}{l|}{\multirow{2}{*}{Betweenness CA}} & \multicolumn{1}{l|}{\multirow{2}{*}{Avg Path Length}} \\
                          & \multicolumn{1}{l|}{}                   & \multicolumn{1}{l|}{}                   & \multicolumn{1}{l|}{}                           & \multicolumn{1}{l|}{}                                  & \multicolumn{1}{l|}{}                              & \multicolumn{1}{l|}{}                                      & \multicolumn{1}{l|}{}                                & \multicolumn{1}{l|}{}                                 \\ \midrule
adjnoun                   & 112                                     & 425                                     & 7,589                                           & 0,072                                                  & 0,403                                              & 0,173                                                      & 37,085                                               & 2,536                                                 \\ \midrule
Dolphins                  & 62                                      & 159                                     & 5,129                                           & 0,091                                                  & 0,307                                              & 0,259                                                      & 39,925                                               & 3,357                                                 \\ \midrule
Football                  & 115                                     & 613                                     & 10,661                                          & 0,092                                                  & 0,399                                              & 0,403                                                      & 26,821                                               & 2,508                                                 \\ \midrule
karate                    & 34                                      & 78                                      & 4,588                                           & 0,146                                                  & 0,426                                              & 0,571                                                      & 17,321                                               & 2,408                                                 \\ \midrule
Lesmis                    & 77                                      & 254                                     & 6,597                                           & 0,078                                                  & 0,389                                              & 0,573                                                      & 30,425                                               & 2,641                                                 \\ \midrule
Polbooks                  & 105                                     & 441                                     & 8,4                                             & 0,072                                                  & 0,33                                               & 0,488                                                      & 38,118                                               & 3,079                                                 \\ \bottomrule
\end{tabular}
\begin{tablenotes}
\item[] Note:  'Deg Avg' to the average degree, ' Eigenvector CA' to the average eigenvector centrality, ' Closeness CA' to the average closeness centrality, ' Clustering Coef. Avg.' to the average clustering coefficient, ' Betweenness CA' to the average betweenness centrality, and ' Avg Path Length ' to the average path length. 
\end{tablenotes}
\end{threeparttable}

\end{table*}

\begin{table*}[ht]
\centering
\caption{Modularity Analysis Using the Girvan-Newman Algorithm}
\begin{threeparttable}

\label{tab:my-table}
\begin{tabular}{@{}|l|ll|ll|ll|ll|@{}}
\toprule
\multirow{2}{*}{Networks} & \multicolumn{2}{l|}{Max modularity}  & \multicolumn{2}{l|}{Elbow diff 2}    & \multicolumn{2}{l|}{Elbow diff 3}    & \multicolumn{2}{l|}{Elbow diff 5}    \\ \cmidrule(l){2-9} 
                          & \multicolumn{1}{l|}{N   Com} & Mod   & \multicolumn{1}{l|}{N   Com} & Mod   & \multicolumn{1}{l|}{N   Com} & Mod   & \multicolumn{1}{l|}{N   Com} & Mod   \\ \midrule
adjnoun                   & \multicolumn{1}{l|}{2}       & 0,009 & \multicolumn{1}{l|}{2}       & 0,009 & \multicolumn{1}{l|}{2}       & 0,009 & \multicolumn{1}{l|}{2}       & 0,009 \\ \midrule
Dolphins                  & \multicolumn{1}{l|}{5}       & 0,519 & \multicolumn{1}{l|}{2}       & 0,381 & \multicolumn{1}{l|}{3}       & 0,381 & \multicolumn{1}{l|}{5}       & 0,519 \\ \midrule
Football                  & \multicolumn{1}{l|}{10}      & 0,6   & \multicolumn{1}{l|}{2}       & 0,4   & \multicolumn{1}{l|}{3}       & 0,455 & \multicolumn{1}{l|}{5}       & 0,55  \\ \midrule
karate                    & \multicolumn{1}{l|}{5}       & 0,401 & \multicolumn{1}{l|}{2}       & 0,36  & \multicolumn{1}{l|}{2}       & 0,36  & \multicolumn{1}{l|}{5}       & 0,401 \\ \midrule
Lesmis                    & \multicolumn{1}{l|}{6}       & 0,459 & \multicolumn{1}{l|}{3}       & 0,26  & \multicolumn{1}{l|}{3}       & 0,26  & \multicolumn{1}{l|}{5}       & 0,26  \\ \midrule
Polbooks                  & \multicolumn{1}{l|}{5}       & 0,517 & \multicolumn{1}{l|}{2}       & 0,443 & \multicolumn{1}{l|}{3}       & 0,483 & \multicolumn{1}{l|}{5}       & 0,517 \\ \bottomrule
\end{tabular}
\begin{tablenotes}
\item[] Note: This same analysis was done for control network 2 based on the Radicchi method (see Table 3) 
\end{tablenotes}
\end{threeparttable}
\end{table*}

\begin{table*}[ht]
\centering
\caption{Modularity Analysis Using the Radicchi Algorithm}
\label{tab:my-table}
\begin{tabular}{@{}|l|ll|ll|ll|ll|@{}}
\toprule
\multirow{2}{*}{Networks} & \multicolumn{2}{l|}{Max modularity}  & \multicolumn{2}{l|}{Elbow diff 2}    & \multicolumn{2}{l|}{Elbow diff 3}    & \multicolumn{2}{l|}{Elbow diff 5}    \\ \cmidrule(l){2-9} 
                          & \multicolumn{1}{l|}{N   Com} & Mod   & \multicolumn{1}{l|}{N   Com} & Mod   & \multicolumn{1}{l|}{N   Com} & Mod   & \multicolumn{1}{l|}{N   Com} & Mod   \\ \midrule
adjnoun                   & \multicolumn{1}{l|}{9}       & 0,175 & \multicolumn{1}{l|}{5}       & 0,13  & \multicolumn{1}{l|}{5}       & 0,13  & \multicolumn{1}{l|}{5}       & 0,13  \\ \midrule
Dolphins                  & \multicolumn{1}{l|}{7}       & 0,467 & \multicolumn{1}{l|}{2}       & 0,257 & \multicolumn{1}{l|}{3}       & 0,263 & \multicolumn{1}{l|}{5}       & 0,31  \\ \midrule
Football                  & \multicolumn{1}{l|}{10}      & 0,585 & \multicolumn{1}{l|}{3}       & 0,286 & \multicolumn{1}{l|}{3}       & 0,286 & \multicolumn{1}{l|}{5}       & 0,459 \\ \midrule
karate                    & \multicolumn{1}{l|}{4}       & 0,373 & \multicolumn{1}{l|}{3}       & 0,373 & \multicolumn{1}{l|}{3}       & 0,373 & \multicolumn{1}{l|}{4}       & 0,373 \\ \midrule
Lesmis                    & \multicolumn{1}{l|}{9}       & 0,515 & \multicolumn{1}{l|}{2}       & 0,373 & \multicolumn{1}{l|}{3}       & 0,481 & \multicolumn{1}{l|}{3}       & 0,481 \\ \midrule
Polbooks                  & \multicolumn{1}{l|}{5}       & 0,521 & \multicolumn{1}{l|}{2}       & 0,457 & \multicolumn{1}{l|}{3}       & 0,484 & \multicolumn{1}{l|}{5}       & 0,521 \\ \bottomrule
\end{tabular}
\end{table*}

Table 1 provides a comprehensive summary of the fundamental characteristics of various networks analyzed in our study. It details crucial aspects such as the number of nodes and edges, which offer an overview of the scale and complexity of each network. Additionally, key centrality metrics such as average degree, betweenness centrality, closeness centrality, along with the clustering coefficient and the average path length are included. These indicators are essential for understanding the internal structure and dynamics of the networks, thus allowing a deeper assessment of their behavior and effectiveness in community detection. Each network, from "adjnoun" to "Polbooks", displays variations in these metrics, reflecting their unique structures and providing valuable insights for future research and applications in computer science and network analysis.
\\
\\
To examine the modularity obtained by the algorithms, we operate on the principle of ignoring the "correct" answer. Instead, we aim to select the community structure that maximizes the basic Girvan algorithm's modularity coefficient. Both Newman and the Radicchi adaptation were tested with 10 communities.
Furthermore, the elbow method, which calculates variations in modularity using a sliding window, was employed to determine the optimal number of communities (see Table 2).
\\
\\
The behavior of modularity and elbow selection can be observed in the context of varying modularity, with comparable outcomes observed between the communities identified through Girvan Newman and those identified through Radicchi.
\begin{figure*}[ht]
    \centering
    \includegraphics[width=0.8\textwidth]{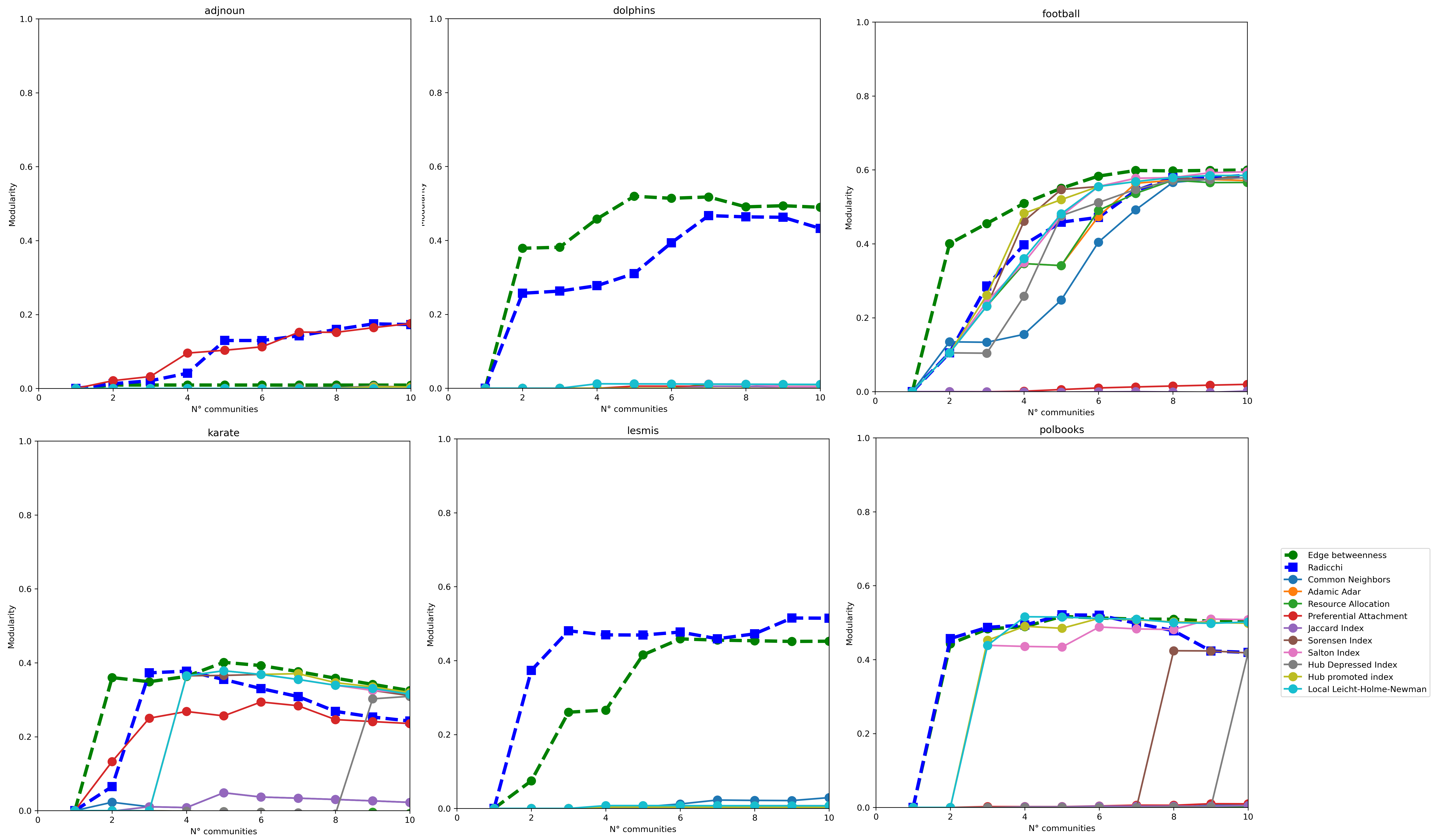}
    \caption{Behavior of Modularity Index Using Local Metrics}
    \label{fig:mesh1}
\end{figure*}

\begin{table}[]
\begin{threeparttable}
\caption{Comparative Analysis of Girvan-Newman and Radicchi Algorithms Using the Elbow Method}
\label{tab:my-table}
\begin{tabular}{@{}|l|ll|ll|ll|ll|@{}}
\toprule
\multirow{2}{*}{Networks} & \multicolumn{2}{l|}{\begin{tabular}[c]{@{}l@{}}Num \\ Communities\end{tabular}} & \multicolumn{2}{l|}{Elbow}    & \multicolumn{2}{l|}{Elbow 3}  & \multicolumn{2}{l|}{Elbow 5}  \\ \cmidrule(l){2-9} 
                          & \multicolumn{1}{l|}{GN}                          & Rad                          & \multicolumn{1}{l|}{GN} & Rad & \multicolumn{1}{l|}{GN} & Rad & \multicolumn{1}{l|}{GN} & Rad \\ \midrule
adjnoun                   & \multicolumn{1}{l|}{2}                           & 9                            & \multicolumn{1}{l|}{2}  & 5   & \multicolumn{1}{l|}{2}  & 5   & \multicolumn{1}{l|}{2}  & 5   \\ \midrule
Dolphins                  & \multicolumn{1}{l|}{5}                           & 7                            & \multicolumn{1}{l|}{2}  & 2   & \multicolumn{1}{l|}{3}  & 3   & \multicolumn{1}{l|}{5}  & 5   \\ \midrule
Football                  & \multicolumn{1}{l|}{10}                          & 10                           & \multicolumn{1}{l|}{2}  & 3   & \multicolumn{1}{l|}{3}  & 3   & \multicolumn{1}{l|}{5}  & 5   \\ \midrule
karate                    & \multicolumn{1}{l|}{5}                           & 4                            & \multicolumn{1}{l|}{2}  & 3   & \multicolumn{1}{l|}{2}  & 3   & \multicolumn{1}{l|}{5}  & 4   \\ \midrule
Lesmis                    & \multicolumn{1}{l|}{6}                           & 9                            & \multicolumn{1}{l|}{3}  & 2   & \multicolumn{1}{l|}{3}  & 3   & \multicolumn{1}{l|}{5}  & 3   \\ \midrule
Polbooks                  & \multicolumn{1}{l|}{5}                           & 5                            & \multicolumn{1}{l|}{2}  & 2   & \multicolumn{1}{l|}{3}  & 3   & \multicolumn{1}{l|}{5}  & 5   \\ \bottomrule
\end{tabular}
\begin{tablenotes}
\item[] Note: 'Rad' refers to  Radicchi,  
\end{tablenotes}
\end{threeparttable}
\end{table}
Table 4 emphasizes the stability and accuracy differences between the Girvan-Newman and Radicchi algorithms by evaluating community structures at 2, 3, and 5 community increments. The analysis identifies the three-term window elbow method as the most stable and effective, particularly when integrated with local metrics to enhance the precision and robustness of community detection.
\\
\\

\subsection{Local Metrics}
The Radicchi algorithm is employed to determine the local evaluation of links, serving as a heuristic to establish their importance within the network and to identify the critical links for partitioning the network. Modularity is evaluated for each configuration, with up to 10 communities detected for each network.
\\
\\
In Figure 1 can be observed that in certain groups, the process of detecting communities based on local metrics exhibits modularity behavior comparable to that detected by the methods of Girvan-Newman and Radicchi. This is exemplified by the football, karate, and polbooks networks. Conversely, the adjnoun, dolphins, and lesmis networks, which employ local metrics, do not effectively build communities that maximize the modularity index in a manner similar to the base methods.
\\
\\
The effectiveness of local analysis techniques varies considerably depending on the characteristics of the analyzed network. Groups where local metrics exhibit some degree of success tend to be better-connected networks, as evidenced by a higher degree and eigenvector centrality average. Conversely, networks where local measures fail to identify communities based on modularity have a lower average clustering coefficient, indicating a less dense network structure.
\\
\\
The karate network demonstrated the greatest efficacy in identifying communities through the application of local metrics to evaluate modularity, as evidenced by the graph above. To assess the degree of concordance between the outcomes yielded by the algorithms of Girvan and Newman, Radicchi, and those derived from local measurements, modularity, and the coefficient of normalized mutual information, a detailed analysis will be conducted.
\\
\\
\begin{table*}[ht]
\centering

\caption{Comparison of Modularity Index and Community Detection using Girvan-Newman, Radicchi, and Local Metrics for the Karate Network}
\label{tab:my-table}
\begin{tabular}{@{}|l|ll|ll|ll|ll|@{}}
\toprule
\multirow{2}{*}{Karate Network} & \multicolumn{2}{l|}{Maximum modularity} & \multicolumn{2}{l|}{Elbow diff 2}   & \multicolumn{2}{l|}{Elbow diff 3}   & \multicolumn{2}{l|}{Elbow diff 5}   \\ \cmidrule(l){2-9} 
                                & \multicolumn{1}{l|}{N Com}    & Mod     & \multicolumn{1}{l|}{N Com} & Mod    & \multicolumn{1}{l|}{N Com} & Mod    & \multicolumn{1}{l|}{N Com} & Mod    \\ \midrule
GN                              & \multicolumn{1}{l|}{5}        & 0,401   & \multicolumn{1}{l|}{2}     & 0,360  & \multicolumn{1}{l|}{2}     & 0,360  & \multicolumn{1}{l|}{5}     & 0,401  \\ \midrule
RAD                             & \multicolumn{1}{l|}{4}        & 0,377   & \multicolumn{1}{l|}{3}     & 0,373  & \multicolumn{1}{l|}{3}     & 0,373  & \multicolumn{1}{l|}{4}     & 0,377  \\ \midrule
CN                              & \multicolumn{1}{l|}{10}       & 0,027   & \multicolumn{1}{l|}{5}     & 0,008  & \multicolumn{1}{l|}{6}     & 0,018  & \multicolumn{1}{l|}{6}     & 0,018  \\ \midrule
AA                              & \multicolumn{1}{l|}{1}        & 0,000   & \multicolumn{1}{l|}{9}     & -0,004 & \multicolumn{1}{l|}{9}     & -0,004 & \multicolumn{1}{l|}{9}     & -0,004 \\ \midrule
RA                              & \multicolumn{1}{l|}{1}        & 0,000   & \multicolumn{1}{l|}{9}     & -0,004 & \multicolumn{1}{l|}{9}     & -0,004 & \multicolumn{1}{l|}{9}     & -0,004 \\ \midrule
PA                              & \multicolumn{1}{l|}{10}       & 0,061   & \multicolumn{1}{l|}{9}     & 0,041  & \multicolumn{1}{l|}{10}    & 0,061  & \multicolumn{1}{l|}{10}    & 0,061  \\ \midrule
JA                              & \multicolumn{1}{l|}{6}        & 0,369   & \multicolumn{1}{l|}{4}     & 0,369  & \multicolumn{1}{l|}{4}     & 0,369  & \multicolumn{1}{l|}{5}     & 0,366  \\ \midrule
SO                              & \multicolumn{1}{l|}{6}        & 0,369   & \multicolumn{1}{l|}{4}     & 0,369  & \multicolumn{1}{l|}{4}     & 0,369  & \multicolumn{1}{l|}{5}     & 0,366  \\ \midrule
SA                              & \multicolumn{1}{l|}{5}        & 0,378   & \multicolumn{1}{l|}{4}     & 0,365  & \multicolumn{1}{l|}{4}     & 0,365  & \multicolumn{1}{l|}{5}     & 0,378  \\ \midrule
HD                              & \multicolumn{1}{l|}{10}       & 0,309   & \multicolumn{1}{l|}{9}     & 0,302  & \multicolumn{1}{l|}{9}     & 0,302  & \multicolumn{1}{l|}{9}     & 0,302  \\ \midrule
HP                              & \multicolumn{1}{l|}{5}        & 0,378   & \multicolumn{1}{l|}{4}     & 0,365  & \multicolumn{1}{l|}{4}     & 0,365  & \multicolumn{1}{l|}{5}     & 0,378  \\ \midrule
LLHN                            & \multicolumn{1}{l|}{5}        & 0,378   & \multicolumn{1}{l|}{4}     & 0,365  & \multicolumn{1}{l|}{4}     & 0,365  & \multicolumn{1}{l|}{5}     & 0,378  \\ \bottomrule
\end{tabular}
\end{table*}
Girvan Newman's basic method is the one that offers a high degree of modularity from the outset of the detection process, reaching a maximum of 5 communities and with the optimal number of communities analysis falling between 2 and 5 communities. Radicchi's method also exhibits a similar behavior, achieving its maximum in 4 communities and with an elbow over 3 and 4 communities, the value of maximum modularity being somewhat lower.
\\
\\
Some metrics that are similar to the results obtained by Girvan Newman are JA (Jaccard), SO (Salton), and SA (Sorensen), which focus on highlighting the similarity between a set of neighboring nodes and achieving modularity results similar to those found by Radicchi. Similarly, methods such as HD (Hub Depressed) and HP (Hub Promoted) focus on the analysis of nodes with a high degree of node connectivity.
\\
\\
In the final stage of the analysis, the most suitable metrics for approximating the behavior of Girvan Newman's base method will be identified for the four community configurations previously identified. 
\\
\\
\begin{table}[]
\caption{Highest Normalized Mutual Information (NMI) Values for the Karate Network Across Different Community Sizes: Comparisons with Girvan-Newman, Radicchi, and High-Performance Local Metrics}
\label{tab:my-table}
\begin{tabular}{@{}|l|r|r|r|r|r|r|@{}}
\toprule
\begin{tabular}[c]{@{}l@{}}Network\\ Karate\end{tabular}  & \multicolumn{1}{l|}{RAD} & \multicolumn{1}{l|}{JA} & \multicolumn{1}{l|}{SO} & \multicolumn{1}{l|}{SA} & \multicolumn{1}{l|}{HP} & \multicolumn{1}{l|}{LLHN} \\ \midrule
                                                          & \multicolumn{1}{l|}{}    & \multicolumn{1}{l|}{}    & \multicolumn{1}{l|}{}    & \multicolumn{1}{l|}{}    & \multicolumn{1}{l|}{}    & \multicolumn{1}{l|}{}      \\ \midrule
\begin{tabular}[c]{@{}l@{}}Communities\\ (2)\end{tabular} & 0,111                    & 0,060                    & 0,060                    & 0,043                    & 0,060                    & 0,060                      \\ \midrule
\begin{tabular}[c]{@{}l@{}}Communities\\ (3)\end{tabular} & 0,671                    & 0,294                    & 0,294                    & 0,294                    & 0,294                    & 0,294                      \\ \midrule
\begin{tabular}[c]{@{}l@{}}Communities\\ (4)\end{tabular} & 0,637                    & 0,707                    & 0,707                    & 0,707                    & 0,707                    & 0,707                      \\ \midrule
\begin{tabular}[c]{@{}l@{}}Communities\\ (5)\end{tabular} & 0,675                    & 0,681                    & 0,681                    & 0,720                    & 0,720                    & 0,720                      \\ \bottomrule
\end{tabular}
\end{table}

\begin{table}[]
\caption{Highest Normalized Mutual Information (NMI) Values for the Karate Network Across Different Community Sizes: Comparisons with Girvan-Newman, Radicchi, and Lower-Performance Local Metrics}
\label{tab:my-table}
\begin{tabular}{@{}lrrrrrr@{}}
\toprule
\begin{tabular}[c]{@{}l@{}}Network\\ Karate\end{tabular}                        & \multicolumn{1}{l}{RAD}    & \multicolumn{1}{l}{CN}    & \multicolumn{1}{l}{AA}    & \multicolumn{1}{l}{RA}    & \multicolumn{1}{l}{PA}    & \multicolumn{1}{l}{HD}             \\ \midrule
\multicolumn{1}{|l|}{}                                                          & \multicolumn{1}{l|}{}      & \multicolumn{1}{l|}{}      & \multicolumn{1}{l|}{}      & \multicolumn{1}{l|}{}      & \multicolumn{1}{l|}{}      & \multicolumn{1}{l|}{}               \\ \midrule
\multicolumn{1}{|l|}{\begin{tabular}[c]{@{}l@{}}Communities\\ (2)\end{tabular}} & \multicolumn{1}{r|}{0,111} & \multicolumn{1}{r|}{0,060} & \multicolumn{1}{r|}{0,043} & \multicolumn{1}{r|}{0,043} & \multicolumn{1}{r|}{0,060} & \multicolumn{1}{r|}{0,043}          \\ \midrule
\multicolumn{1}{|l|}{\begin{tabular}[c]{@{}l@{}}Communities\\ (3)\end{tabular}} & \multicolumn{1}{r|}{0,671} & \multicolumn{1}{r|}{0,294} & \multicolumn{1}{r|}{0,294} & \multicolumn{1}{r|}{0,294} & \multicolumn{1}{r|}{0,080} & \multicolumn{1}{r|}{0,294}          \\ \midrule
\multicolumn{1}{|l|}{\begin{tabular}[c]{@{}l@{}}Communities\\ (4)\end{tabular}} & \multicolumn{1}{r|}{0,637} & \multicolumn{1}{r|}{0,306} & \multicolumn{1}{r|}{0,281} & \multicolumn{1}{r|}{0,281} & \multicolumn{1}{r|}{0,130} & \multicolumn{1}{r|}{\textbf{0,252}} \\ \midrule
\begin{tabular}[c]{@{}l@{}}Communities\\ (5)\end{tabular}                       & 0,675                      & 0,298                      & 0,243                      & 0,243                      & 0,239                      & 0,243                               \\ \bottomrule
\end{tabular}
\end{table}
For the analyzed network, the NMI (Normalized Mutual Information) was high in the local metrics when they exceeded three communities, achieving a rate of approximately 70\% or more in the allocation of nodes to the same communities detected by the Girvan-Newman base method, even surpassing Radicchi's method by a small margin.
\\
\\
Regarding the metrics that yielded unsatisfactory results, it should be noted that they are not entirely without merit. In most cases, they yielded indicators above 24\%, indicating a tendency to detect communities in a manner similar to the original method.
\\
\\
 \newpage
\section{conclusions}

In this paper, we address the problem of community detection by exploiting the intrinsic properties of the network and using local link prediction methods within a hierarchical framework. This framework integrates the core features of the Girvan-Newman and Radicchi algorithms with modularity metrics and normalized mutual information to evaluate the performance of different local metrics.
\\
\\
The basic algorithm used is the one developed by Girvan and Newman, which uses betweenness metrics as cutoff and detection criterion and gives excellent results. However, its computational complexity makes it impractical for large networks. To solve this problem, Radicchi's algorithm was developed, which uses the clustering coefficient as the clustering criterion and achieves good results. However, discrepancies in the detection of the optimal number of communities between the two methods suggest a need for further research to develop heuristics that can improve the process.
\\
\\
We investigated local metrics for community detection. Initial measurements showed variable results depending on network characteristics. Techniques using local metrics showed similar results to the basic algorithms in some networks. However, their effectiveness was significantly limited in less dense networks. It is possible to identify local metrics that offer promising results, matching or even exceeding the performance of Radicchi's algorithm.
\\
\\
To confirm this behavior, the Karate network was used as a reference, and the performance was evaluated using the modularity coefficient and the NMI. This approach aimed to evaluate the performance in terms of node density and compare it with the results obtained by the Girvan-Newman basic algorithm. The results were promising in terms of both modularity and NMI over various local metrics.
\\
\\
Future work will entail a more comprehensive analysis with the objective of identifying characteristics that can enhance local technique algorithms. Additionally, preprocessing of networks will be performed with the aim of highlighting detection characteristics and those that can impact performance.

\bibliographystyle{ieeetr}
\bibliography{citas.bib}

\begin{thebibliography}{10}

\bibitem{1}
S.~Fortunato, ``Community detection in graphs,'' {\em Physics Reports},
  vol.~486, pp.~75--174, Feb. 2010.

\bibitem{2}
W.~Liu, M.~Chen, H.~Miao, Y.~Zhou, and X.~Chen, ``A fast community detection
  algorithm,'' in {\em 2016 5th {International} {Conference} on {Computer}
  {Science} and {Network} {Technology} ({ICCSNT})}, pp.~345--349, Dec. 2016.

\bibitem{3}
M.~Girvan and M.~E.~J. Newman, ``Community structure in social and biological
  networks,'' {\em Proceedings of the National Academy of Sciences}, vol.~99,
  pp.~7821--7826, June 2002.

\bibitem{4}
F.~Radicchi, C.~Castellano, F.~Cecconi, V.~Loreto, and D.~Parisi, ``Defining
  and identifying communities in networks,'' {\em Proceedings of the National
  Academy of Sciences}, vol.~101, pp.~2658--2663, Mar. 2004.

\bibitem{5}
V.~Martínez, F.~Berzal, and J.-C. Cubero, ``A {Survey} of {Link} {Prediction}
  in {Complex} {Networks},'' {\em ACM Computing Surveys}, vol.~49, pp.~1--33,
  Dec. 2016.

\bibitem{6}
D.~Easley and J.~Kleinberg, ``Networks, {Crowds}, and {Markets}:,''

\bibitem{7}
K.~H. Rosen, {\em Discrete mathematics and its applications}.
\newblock New York: McGraw-Hill, 7th ed~ed., 2012.

\bibitem{8}
M.~E.~J. Newman and M.~Girvan, ``Finding and evaluating community structure in
  networks,'' {\em Physical Review E}, vol.~69, p.~026113, Feb. 2004.

\bibitem{9}
M.~E.~J. Newman, ``Detecting community structure in networks,'' {\em The
  European Physical Journal B - Condensed Matter}, vol.~38, pp.~321--330, Mar.
  2004.

\bibitem{10}
{\em Structure in {Complex} {Networks}}, vol.~766 of {\em Lecture {Notes} in
  {Physics}}.
\newblock Berlin, Heidelberg: Springer Berlin Heidelberg, 2009.
\newblock ISSN: 0075-8450.

\bibitem{11}
M.~E.~J. Newman, ``The structure and function of complex networks,'' {\em SIAM
  Review}, vol.~45, pp.~167--256, Jan. 2003.
\newblock arXiv: cond-mat/0303516.

\bibitem{12}
X.~Pan, G.~Xu, B.~Wang, and T.~Zhang, ``A {Novel} {Community} {Detection}
  {Algorithm} {Based} on {Local} {Similarity} of {Clustering} {Coefficient} in
  {Social} {Networks},'' {\em IEEE Access}, vol.~7, pp.~121586--121598, 2019.
\newblock Conference Name: IEEE Access.

\bibitem{13}
D.~Liben-Nowell and J.~Kleinberg, ``The {Link}-{Prediction} {Problem} for
  {Social} {Networks},'' p.~23.

\bibitem{14}
T.~Zhou, L.~Lü, and Y.-C. Zhang, ``Predicting missing links via local
  information,'' {\em The European Physical Journal B}, vol.~71, pp.~623--630,
  Oct. 2009.

\bibitem{15}
J.~C. Valverde-Rebaza and A.~de~Andrade~Lopes, ``Link {Prediction} in {Complex}
  {Networks} {Based} on {Cluster} {Information},'' in {\em Advances in
  {Artificial} {Intelligence} - {SBIA} 2012} (L.~N. Barros, M.~Finger, A.~T.
  Pozo, G.~A. Gimenénez-Lugo, and M.~Castilho, eds.), Lecture {Notes} in
  {Computer} {Science}, pp.~92--101, Springer Berlin Heidelberg, 2012.

\bibitem{16}
Z.~Wu, Y.~Lin, J.~Wang, and S.~Gregory, ``Link prediction with node clustering
  coefficient,'' {\em Physica A: Statistical Mechanics and its Applications},
  vol.~452, pp.~1--8, June 2016.

\bibitem{17}
A.~Kumar, S.~S. Singh, K.~Singh, and B.~Biswas, ``Level-2 node clustering
  coefficient-based link prediction,'' {\em Applied Intelligence}, vol.~49,
  pp.~2762--2779, July 2019.

\bibitem{18}
L.~Lü and T.~Zhou, ``Link prediction in complex networks: {A} survey,'' {\em
  Physica A: Statistical Mechanics and its Applications}, vol.~390,
  pp.~1150--1170, Mar. 2011.

\bibitem{19}
L.~A. Adamic and E.~Adar, ``Friends and neighbors on the {Web},'' {\em Social
  Networks}, vol.~25, pp.~211--230, July 2003.

\bibitem{20}
A.-L. Barabási and R.~Albert, ``Emergence of {Scaling} in {Random}
  {Networks},'' {\em Science}, vol.~286, pp.~509--512, Oct. 1999.

\bibitem{21}
``Étude de la distribution florale dans une portion des {Alpes} et du
  {Jura},'' vol.~37, no.~142.

\bibitem{22}
T.~Sørensen, {\em A method for establishing groups of equal amplitude in plant
  sociology based on similarity of species content and its application to
  analyses of the vegetation on {Danish} commons}, vol.~V of {\em Det
  {Kongelige} {Danske} {Videnskabernes} {Selskab} {Biologiske} {Skrifter}}.
\newblock Denmark: Bianco Lunos Bogtrykkeeri, 1948.

\bibitem{23}
G.~Salton and M.~J. McGill, {\em Introduction to {Modern} {Information}
  {Retrieval}}.
\newblock New York, NY, USA: McGraw-Hill, Inc., 1986.

\bibitem{24}
A.~Rodriguez, B.~Kim, M.~Turkoz, J.-M. Lee, B.-Y. Coh, and M.~K. Jeong, ``New
  {Multi}-stage {Similarity} {Measure} for {Calculation} of {Pairwise} {Patent}
  {Similarity} in a {Patent} {Citation} {Network},'' {\em Scientometrics},
  vol.~103, pp.~565--581, May 2015.

\bibitem{25}
E.~Ravasz, A.~L. Somera, D.~A. Mongru, Z.~N. Oltvai, and A.-L. Barabási,
  ``Hierarchical {Organization} of {Modularity} in {Metabolic} {Networks},''
  {\em Science}, vol.~297, pp.~1551--1555, Aug. 2002.

\bibitem{26}
E.~A. Leicht, P.~Holme, and M.~E.~J. Newman, ``Vertex similarity in networks,''
  {\em Physical Review E}, vol.~73, p.~026120, Feb. 2006.
\newblock arXiv: physics/0510143.

\bibitem{27}
M.~Newman, {\em Networks: {An} {Introduction}}.
\newblock Oxford University Press, Mar. 2010.

\bibitem{28}
L.~Ana and A.~Jain, ``Robust data clustering,'' in {\em 2003 {IEEE} {Computer}
  {Society} {Conference} on {Computer} {Vision} and {Pattern} {Recognition},
  2003. {Proceedings}.}, vol.~2, (Madison, WI, USA), pp.~II--128--II--133, IEEE
  Comput. Soc, 2003.

\bibitem{29}
S.~Wagner and D.~Wagner, ``Comparing {Clusterings} - {An} {Overview},'' p.~19.

\bibitem{30}
R.~A. Rossi and N.~K. Ahmed, ``The {Network} {Data} {Repository} with
  {Interactive} {Graph} {Analytics} and {Visualization},'' in {\em {AAAI}},
  2015.

\bibitem{31}
J.~Leskovec and A.~Krevl, ``{SNAP} {Datasets}: {Stanford} {Large} {Network}
  {Dataset} {Collection},'' June 2014.

\bibitem{32}
V.~Batagelj and A.~Mrvar, ``Pajek datasets,'' 2006.

\bibitem{33}
W.~W. Zachary, ``An {Information} {Flow} {Model} for {Conflict} and {Fission}
  in {Small} {Groups},'' {\em Journal of Anthropological Research}, vol.~33,
  no.~4, pp.~452--473, 1977.

\bibitem{34}
D.~Lusseau, K.~Schneider, O.~J. Boisseau, P.~Haase, E.~Slooten, and S.~M.
  Dawson, ``The bottlenose dolphin community of {Doubtful} {Sound} features a
  large proportion of long-lasting associations,'' {\em Behavioral Ecology and
  Sociobiology}, vol.~54, pp.~396--405, Sept. 2003.

\bibitem{35}
D.~E. Knuth, {\em The {Stanford} {GraphBase}: {A} {Platform} for
  {Combinatorial} {Computing}}.
\newblock New York, NY, USA: ACM, 1993.

\bibitem{36}
V.~Krebs, ``Political {Polarization} {During} the 2008 {US} {Presidential}
  {Campaign},'' 2008.

\bibitem{37}
M.~E.~J. Newman, ``Modularity and community structure in networks,'' {\em
  Proceedings of the National Academy of Sciences}, vol.~103, p.~8577, June
  2006.

\end{thebibliography}

\begin{IEEEbiography}[{\includegraphics[width=1in,height=1.25in,clip,keepaspectratio]{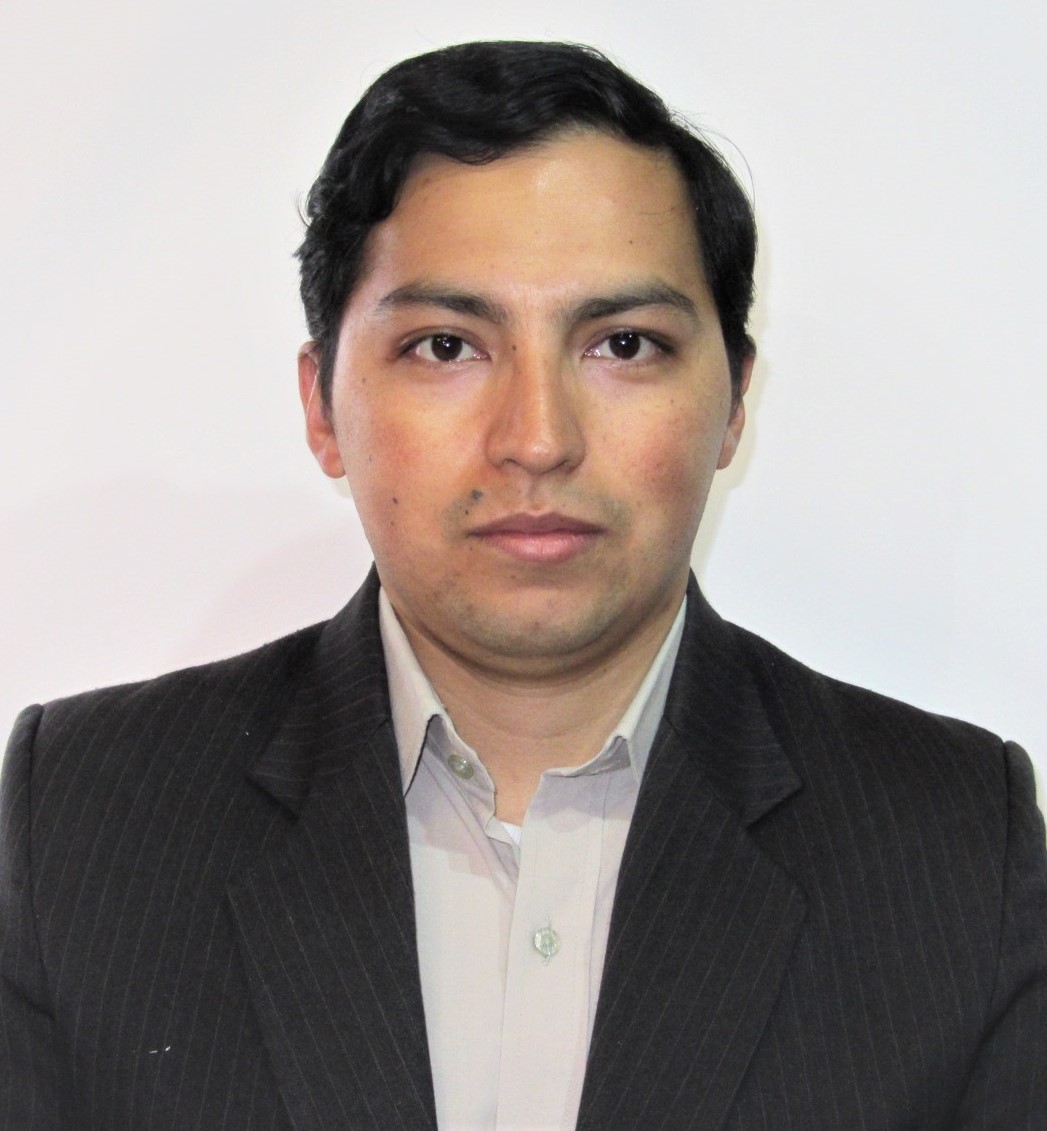}}]{Julio-Omar Palacio-Niño} is a Ph.D. student in Information and Communication Technologies in the Department of Computer Science and Artificial Intelligence at the University of Granada, Spain. He received a B.E. degree in System Engineering from the National University of Colombia, in 2009, M.Sc. degree in Telecommunication Engineering from the Nacional University of Colombia, in 2012 and M.Sc. degree in Soft Computing and Artificial Intelligence from the University of Granada, in 2013. Since 2021, he has been an Assistant Professor with the Systems Engineering Department, Pontificia Universidad Javeriana. His research interests include networks analysis, data mining, machine learning and deep learning.
\end{IEEEbiography}	

\begin{IEEEbiography}[{\includegraphics[width=1in,height=1.25in,clip,keepaspectratio]{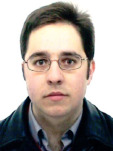}}]{Fernando Berzal} is an associate professor in the Department of Computer Science and Artificial Intelligence at the University of Granada. Previously, he had been a visiting research scientist at the data mining research group led by Jiawei Han at the University of Illinois at Urbana-Champaign. His research interests include model-driven software development, software design, and the application of data mining techniques to software engineering problems. He received his Ph.D. in Computer Science from the University of Granada in 2002 and he was awarded the Computer Science Studies National First Prize by the Spanish Ministry of Education in 2000. He is a senior member of the ACM and also a member of the IEEE Computer Society. 
\end{IEEEbiography}

\end{document}